\documentclass[11pt,osid]{article}

\usepackage{osid}
\usepackage{amsfonts}
\usepackage{amsmath}
\usepackage{graphicx}

\def\ket#1{|#1\rangle}

\def\ketbra#1#2{|#1\rangle\!\langle#2|}
\def\braket#1#2{\langle#1|#2\rangle}

\newcommand{\id}{\mathbb{I}}
\newcommand{\im}{\mathrm{i}}

\title{Efficiency in Quantum Key Distribution protocols
with entangled gaussian states}
\author{C. Rod{\'o}\\
    {\footnotesize\it Grup de F{\'\i}sica Te{\`o}rica, Universitat Aut{\`o}noma de Barcelona, 08193 Spain \& rodo@ifae.es}\\[2ex]
    O. Romero-Isart\\
    {\footnotesize\it Grup de F{\'\i}sica Te{\`o}rica, Universitat Aut{\`o}noma de Barcelona, 08193 Spain \& ori@ifae.es}\\[2ex]
    K. Eckert\\
    {\footnotesize\it Grup de F{\'\i}sica Te{\`o}rica, Universitat Aut{\`o}noma de Barcelona, 08193 Spain \& kai@ifae.es}\\[2ex]
    A. Sanpera\\
    {\footnotesize\it ICREA and Grup de F{\'\i}sica Te{\`o}rica, Universitat Aut{\`o}noma de Barcelona, 08193 Spain \& sanpera@ifae.es}}

\begin{document}

\maketitle

\begin{abstract}
    Quantum key distribution (QKD) refers to specific quantum
    strategies which permit the secure distribution of a secret key
    between two parties that wish to communicate secretly. Quantum
    cryptography has proven unconditionally secure in ideal
    scenarios and has been successfully implemented using quantum
    states with finite (discrete) as well as infinite (continuous)
    degrees of freedom. Here, we analyze the efficiency of QKD
    protocols that use as a resource entangled gaussian states and
    gaussian operations only. In this framework, it has already
    been shown that QKD is possible \cite{Navascues05} but the
    issue of its efficiency has not been considered. We propose a
    figure of merit (the efficiency $E$) to quantify the number of
    classical correlated bits that can be used to distill a key
    from a sample of $N$ entangled states. We relate the
    efficiency of the protocol to the entanglement and purity of
    the states shared between the parties.
\end{abstract}

\section{Introduction}

Quantum cryptography relies on the possibility of establishing a
secret random key between two distant parties traditionally
denoted as Alice and Bob. If the key is securely distributed, the
algorithms used to encode and decode any message can be made
public without compromising security. The key consists typically
in a random sequence of bits which both, Alice and Bob, share as a
string of classically correlated data. The superiority of quantum
cryptography comes from the fact that the laws of quantum
mechanics permit to the legitimate users (Alice and Bob) to infer
if an eavesdropper has monitored the distribution of the key and
has gained information about it. If this is the case, Alice and
Bob will both agree in withdrawing the key and will start the
distribution of a new one. In contrast, classical key
distribution, no matter how difficult the distribution from a
technological point of view is, can always be intercepted by an
eavesdropper without Alice and Bob realizing it. In quantum
cryptography, there exist several protocols that Alice and Bob can
use in order to establish a secret key. Some of them, like Ekert91
\cite{Ekert91}, use as a resource shared entanglement between the
two parties, while in others, like BB84 \cite{BB84}, the key is
established by sending non entangled quantum states between the
parties and communicating classically. If Alice and Bob share a
collection of distillable entangled states, they can always obtain
from them a smaller number of maximally entangled states from
which they can establish a secure key \cite{Deutsch96}. The number
of singlets (maximally entangled states) that can be extracted
from a quantum state using only Local Operations and Classical
Communication (LOCC) is referred to as the Entanglement of
Distillation $E_D$. For establishing a key, another important
concept is the number of secret bits $K_D$, that can be extracted
from a quantum state using LOCC. Since a secret bit can always be
extracted from a maximally entangled state, $E_D \leq K_D$. There
exist also quantum states which are entangled but cannot be
distilled, {\it i.e.}, have $E_D=0$. They are usually referred to
as bound entangled states since its entanglement is bound to the
state. Nevertheless, for some of those states it has been shown
that $K_D \neq 0$, and thus, they can be used to establish a
secret key \cite{Horodecki05}.

A particular case of states that cannot be ``distilled'' by
``normal'' procedures are continuous variables gaussian states,
{\it e.g.}, thermal, coherent, and squeezed states of light. By
``normal'' procedures we mean operations that preserve the
gaussian character of the state (gaussian operations). They
correspond {\it e.g.}, to beam splitters, squeezers, mirrors, etc.
Thus, in the gaussian scenario all entangled gaussian states
posses bound entanglement. Quantum cryptography with gaussian
states using gaussian operations has been experimentally
implemented using ``prepare and measure schemes'' with either
squeezed or coherent states \cite{Preskill01, Grangier02}. Those
schemes do not demand entanglement between the parties.

Recently, Navascu{\'e}s {\it et al.} \cite{Navascues05} have shown
that it is also possible with only gaussian operations to extract
a secret key {\it {\`a} la} Ekert91 from entangled gaussian states, in
spite the fact that these states are not gaussian distillable. In
other words, it has been shown that in the gaussian scenario all
entangled gaussian states fulfill $GK_D>0$ (where the letter $G$
stands for gaussian) while $GE_D=0$. Alice and Bob can extract a
list of classically correlated bits from a set a of $1 \times 1$
entangled modes as follows: i) they agree on a value $x_0 > 0$,
ii) Alice (Bob) measures the quadrature of each of her (his) modes
$\hat X_A (\hat X_B)$, iii) they accept only outputs such that
$|x_A|=|x_B|=x_0$, iv) they associate {\it e.g.}, the classical
value $0$(1) to $x_i=+x_0(-x_0)$, $i=A, B$ and thus establish a
list of classically correlated bits. From there, they can apply
Advantage Distillation \cite{Maurer93} to establish the secret
key. This protocol is secure against individual eavesdropper
attacks. Since the protocol is based on output coincidences of the
measurements of the quadratures which, by definition, are
operators with a continuous spectrum, the protocol has zero
efficiency \cite{Navascues05}.

Here, we study the consequences of relaxing the above condition to
a more realistic scenario. We assume that Alice and Bob can
extract a list of sufficiently correlated classical bits obtained
by accepting measurement outputs that do not coincide but are
bound within a range. We ask ourselves which is the possibility
that Alice and Bob can still distribute the key in a secure way
under individual and coherent attacks. We obtain that there exists
always a finite interval for which the protocol can be implemented
successfully. The length of this interval depends on the
entanglement and on the purity of the shared states, and increases
with increasing entanglement.

The paper is organised as follows. In Sect. $2$, we present the
formalism needed to tackle this problem. In Sect. $3$, we first
review the previous protocol \cite{Navascues05} and present our
new results. Finally, we present our conclusions in Sect. $4$.

\section{Formalism}\label{Formalism}

Systems of continuous variables are often expressed in terms of
modes, where each mode has two associated canonical degrees of
freedom (``position'' and ``momentum'') which fulfill the
canonical commutation relations (CCR). The CCR for a quantum
system with $n$ modes can be compactly expressed {\it via} the
symplectic matrix. Denoting the canonical coordinates by
\begin{equation}
    \hat{\mathbf{R}}^T =
    (\hat{q}_1,\hat{p}_1,\ldots,\hat{q}_n,\hat{p}_n) \equiv
    (\hat{R}_1,\ldots,\hat{R}_{2n}),\nonumber
\end{equation}
the CCR simply read $[\hat{R}_i,\hat{R}_j] = \im(J_n)_{ij}$, where
$i,j=1,\ldots,2n$ and
\begin{equation}
    \mathbf{J}_n = \bigoplus_{i=1}^n \mathbf{J}, \quad \quad
    \mathbf{J} \equiv
    \begin{pmatrix}
     0 & 1\\
     -1 & 0
    \end{pmatrix}.
\end{equation}
The symplectic matrix $\mathbf{J}$ defines the symplectic scalar
product and describes the geometry of the phase space. There is a
bijective map between a quantum state described by a density
matrix $\hat \rho$ (in an infinite Hilbert space) and its
corresponding characteristic function $\chi_{\rho}$, which is
given by the Fourier-Weyl transform:
\begin{equation}\label{Characteristic}
    \chi_{\rho}(\boldsymbol{\zeta}) \equiv \rm{tr} \{\hat{\rho}
    \hat{W}_{(\boldsymbol{\zeta})} \},
\end{equation}
\begin{equation}
    \hat{\rho} \equiv \frac{1}{(2 \pi)^n} \int d^{2n}
    \boldsymbol{\zeta} \, \chi_{\rho}(\boldsymbol{\zeta})
    \hat{W}_{(-\boldsymbol{\zeta})},
\end{equation}
where $\boldsymbol{\zeta} \in \mathbb{R}^{2n}$ and
$\hat{W}_{\boldsymbol{\zeta}} = e^{\im
\boldsymbol{\boldsymbol{\zeta}}^T \mathbf{J}_n
\hat{\boldsymbol{R}}}$ are the so-called Weyl operators. Gaussian
states are characterized by a gaussian $\chi_\rho$ function,
\begin{equation}\label{Characteristic2}
    \chi_\rho(\boldsymbol{\zeta}) = e^{\im \boldsymbol{\zeta}^T
    \cdot \mathbf{J}_n \cdot \mathbf{d} - \frac{1}{4}
    \boldsymbol{\zeta}^T \mathbf{J}^T_n \cdot \boldsymbol{\gamma}
    \cdot \mathbf{J}_n \boldsymbol{\zeta}},
\end{equation}
where $\mathbf{d}$ is a $2n$ real vector, called displacement
vector (DV), and $\boldsymbol{\gamma}$ is a $2n \times 2n$
symmetric real matrix, denoted as covariance matrix (CM). A
convenient representation of gaussian quantum states is given in
terms of the Wigner quasi-distribution function $\mathcal{W}_\rho$
\cite{Wigner32}, which is related to the characteristic function
by the symplectic Fourier transform which preserves the gaussian
character,
\begin{eqnarray}
    \mathcal{W}_{\rho}(\boldsymbol{\zeta})& =& \frac{1}{(2 \pi)^{2n}}
    \int d^{2n} \boldsymbol{\eta} \, \chi_{\rho}(\boldsymbol{\eta})
    e^{-\im \boldsymbol{\eta}^T \cdot \mathbf{J}_n \cdot
    \boldsymbol{\zeta}}, \\
     \chi_{\rho}(\boldsymbol{\eta}) &=&
    \int d^{2n} \boldsymbol{\zeta}
    \mathcal{W}_{\rho}(\boldsymbol{\zeta}) e^{\im \boldsymbol{\eta}^T
    \cdot \mathbf{J}_n \cdot \boldsymbol{\zeta}},
\end{eqnarray}
where $\boldsymbol{\eta} \in \mathbb{R}^{2n}$. Thus, a gaussian
quantum state can equivalently be defined as a quantum state whose
Wigner function is gaussian,
\begin{equation}
    \mathcal{W}_\rho(\boldsymbol{\zeta}) = \frac{1}{\pi^n \sqrt{{\rm
    det} \boldsymbol{\gamma}}} e^{-(\boldsymbol{\zeta}-\mathbf{d)^T
    \cdot \frac{1}{\boldsymbol{\gamma}} \cdot (\boldsymbol{\zeta}
    -\mathbf{d})}}.
\end{equation}
$\mathbf{d}$ and $\boldsymbol{\gamma}$ are defined as:
\begin{equation}
    d_i = {\rm tr} (\hat{\rho} \hat{R}_i),
\end{equation}
\begin{equation}
    \gamma_{ij} = {\rm tr} (\hat{\rho} \{ \hat{R}_i-d_i \hat{\id},
    \hat{R}_j-d_j \hat{\id} \}),
\end{equation}
and are computed {\it via} the first and second moments of the
characteristic function,
\begin{equation}
    \left. d_i' = -\im \frac{\partial}{\partial \zeta_i} \chi_{\rho}
    (\boldsymbol{\zeta}) \right|_{\boldsymbol{\zeta} = 0} = {\rm
    tr} (\hat{\rho} \hat{R'}_i),
\end{equation}
\begin{equation}
    \left. \frac{\gamma'_{ij}}{2} + d_i'd_j' = (-\im)^2
    \frac{\partial^2}{\partial \zeta_i \partial \zeta_j} \chi_{\rho}
    (\boldsymbol{\zeta}) \right|_{\boldsymbol{\zeta} = 0} = \frac{1}
    {2}{\rm tr}(\hat{\rho} \{ \hat{R'}_i, \hat{R'}_j \}),
\end{equation}
where  $\hat{R}'_i = J_{ij} \hat{R}_j$, $d'_i = J_{ij} d_j$ and
$\gamma'_{ij} = J^T_{ik} \gamma_{kl} J_{lj}$.

In analogy with classical probability theory, the displacement
vector $\mathbf{d}$ plays the role of the mean value $\mu_i = {\rm
E}[x_i]$, and the covariance matrix elements $\boldsymbol{\gamma}$
play the role of the covariances $C_{ij} = {\rm Cov}(x_i,x_j) =
{\rm E}[(x_i-\mu_i)(x_j-\mu_j)]$ of a classical probability
distribution. So only relative displacement vectors have physical
meaning and only the non block-diagonal terms of the covariance
matrix tell us about the quantum correlations present in the
state.

Since the density matrix is a semidefinite positive operator,
$\hat\rho \geq 0$, the corresponding covariance matrix must
fulfill: $\boldsymbol{\gamma} + \im \mathbf{J}_n \geq 0$. One can
also define the fidelity between continuous gaussian states in
terms of Wigner functions. We use here the Bures-Uhlmann fidelity
between two arbitrary states $\hat\rho_1$ and $\hat\rho_2$ defined
as \cite{Barnum96}
\begin{equation}
    \mathcal{F}(\hat{\rho}_1, \hat{\rho}_2) = \left [ {\rm tr}
    \sqrt{\sqrt{\hat{\rho}_1} \hat{\rho}_2 \sqrt{\hat{\rho}_1}} \right
    ]^2,\nonumber
\end{equation}
which coincides with the so called Hilbert-Schmidt fidelity
\begin{equation}
    \mathcal{F}(\hat{\rho}_1, \hat{\rho}_2) = {\rm tr} (\hat{\rho}_1
    \hat{\rho}_2),
\end{equation}
whenever at least one of the states is pure. At the level of CM,
using the Quantum Parseval relation \cite{Holevo82}, the
Hilbert-Schmidt fidelity between two gaussian states can be
written as:
\begin{equation}
    \begin{split}
     \mathcal{F} (\hat{\rho}_1, \hat{\rho}_2) &= (2\pi)^{n} \int
     d^{2n} \boldsymbol{\zeta} \, \mathcal{W}_1( \boldsymbol{\zeta})
     \mathcal{W}_2 ( \boldsymbol{\zeta}) =\\
     &= \frac{1}{\sqrt{{\rm det} \left( \frac{\boldsymbol{\gamma}_1 +
     \boldsymbol{\gamma}_2}{2} \right) }} e^{-(\mathbf{d}_2 -
     \mathbf{d}_1)^T \cdot \left( \frac{1}{\boldsymbol{\gamma}_1 +
     \boldsymbol{\gamma}_2} \right) \cdot(\mathbf{d}_2-\mathbf{d}_1)}
    \end{split}
\end{equation}
where $\boldsymbol{\gamma}_{1(2)}$ and $\mathbf{d}_{1(2)}$ belong
to $\hat{\rho}_{1(2)}$. Clearly, only relative DVs are of physical
significance. The purity of the state translates to
$\mathcal{P}(\boldsymbol{\gamma}) = {\rm tr}(\hat{\rho}^2) = {\rm
det}(\boldsymbol{\gamma})^{-1/2} \leq 1$. It is important to
notice that gaussian states always admit a purification. Thus, any
mixed gaussian state of $n$ modes can be expressed as the
reduction of a pure gaussian state of $2n$ modes of the form:
\begin{equation}\nonumber
    \boldsymbol{\gamma}_{2n} =
    \begin{pmatrix}
     \boldsymbol{\gamma}_{n} & \mathbf{C}_n\\
     \mathbf{C}_n^T & \boldsymbol{\theta}_n \boldsymbol{\gamma}_{n}
     \boldsymbol{\theta}_n^T
    \end{pmatrix}, \quad \quad
    \mathbf{C}_n = \boldsymbol{J}_n \sqrt{-(\boldsymbol{J}_n
    \boldsymbol{\gamma}_n)^2-\id} \, \boldsymbol{\theta}_n, \quad
    \quad \boldsymbol{\theta}_n = \bigoplus_{i=1}^n
    \boldsymbol{\theta},
\end{equation}
such that the mixed state can be obtained after tracing out $n$
modes from $\boldsymbol{\gamma}_{2n}$. Here $ \boldsymbol{\theta}
= \bigl(
\begin{smallmatrix}
    1 & 0\\
    0 & -1
\end{smallmatrix}
\bigr)$, which is the momentum reflection in phase-space, is the
associated symplectic operation.

For what follows it is also important to study entanglement
properties in the formalism of covariance matrices. A necessary
and sufficient condition for separability for an arbitrary
bipartite state is given in \cite{Giedke01}. For the case $1
\times 1$ and $1 \times N$ modes a necessary and sufficient
condition for separability is provided by the PPT criterion
\cite{Werner01} (for the rest of the states this criterion is only
necessary but not sufficient). The PPT criterion tells us that a
state is entangled if and only if the state $\hat{\rho}$ has non
positive partial transposition (NPPT): $\hat{\rho}^{T_A} < 0$. In
terms of CMs this criterion reads  $ \boldsymbol{\theta}_A
\boldsymbol {\gamma}\boldsymbol{\theta}_A^T + \im \boldsymbol{J} <
0$. The second property, particularly relevant in what follows, is
that any NPPT gaussian state can be mapped by Gaussian Local
Operations and Classical Communication (GLOCC) to an NPPT
symmetric state of $1 \times 1$ modes.

To quantify the entanglement of our states, we will use the
logarithmic negativity as entanglement measure; ${\rm LN}
(\hat\rho) = \log_2 ||\hat\rho^{T_A}||_1$, where $||\hat A||_1 =
{\rm tr} \sqrt{\hat A^\dag \hat A}$ can  be easily computed
through the sum of the singular values of $\hat A$. One can extend
this measure to gaussian states of $n$ modes through
\cite{Plenio07}:
\begin{equation}
    {\rm LN} (\boldsymbol{\gamma}_n) = -\sum_{i=1}^n\log_2
    {\rm min}(\tilde{\mu}_i,1)
\end{equation}
where $\{ \pm\tilde{\mu}_i \} = {\rm spec}(-i \boldsymbol{J}_n
\boldsymbol{\gamma}_n^{T_A})$ {\it i.e.}, the symplectic spectrum
of the partial transposed CM.

With this formalism at hand we now move to
the presentation of our calculations and results.

\section{Results}\label{Results}

First, we summarize the main steps of the protocol used in
\cite{Navascues05}. Without loosing generality, and by virtue of
the properties of gaussian states, one should only consider the
case in which Alice and Bob share many copies of a quantum system
of $1 \times 1$ symmetric NPPT gaussian state $\hat \rho_{AB}$. To
extract a list of classically correlated bits to establish a
secret key, each party measures the quadratures of her/his mode
$\hat X_{A,B}$ and accepts only those outputs $x_{A,B}$ for which
both parties have a consistent result  $|x_A|=|x_B|=x_0$. With
probability $p(i,j)$, each party associates the
classical bit $i=0(1)$ to her/his outcome $+x_0(-x_0)$. The
probability that their symbols do not coincide is given by
$\epsilon_{AB}=(\sum_{i\neq j}p(i,j)) / (\sum_{i,j}p(i,j))$.
Having fixed a string of $N$ classical correlated values, they can
apply classical advantage distillation \cite{Maurer93}. To this
aim, Alice generates a random bit $b$ and encodes her string of $N$
classical bits into a vector $\vec b$ of length $N$ such that
$b_{Ai}+b_i=b \mbox{ mod }2$. Bob checks that for his symbols all
results $b_{Bi}+b_i=b' \mbox{ mod }2$ are consistent, and in this
case accepts the bit $b$. The new error probability is given by
\begin{equation}
    \epsilon_{AB,N} = \frac{(\epsilon_{AB})^N}{(1 - \epsilon_{AB}
    )^N+(\epsilon_{AB})^N} \leq \left( \frac{\epsilon_{AB}}{1 -
    \epsilon_{AB}} \right)^N,
\end{equation}
which tends to zero for sufficiently large $N$. The most general
scenario for eavesdropping is to assume that Eve has access to the
states before their distribution. Hence, the states that Alice and
Bob share correspond to the reduction of a pure 4-mode state. Now
security with respect to individual attacks from the eavesdropper
Eve, can be established if
\begin{equation}
    \left(\frac{\epsilon_{AB}}{1 - \epsilon_{AB}}\right)^N <
    |\braket{e_{++}}{e_{--}}|^N,
\end{equation}
where $\ket{e_{\pm \pm}}$ denotes the state of Eve once Alice and
Bob have projected their states onto $\ket{\pm x_0}$. Notice that
Eve can gain information if the overlap between her states after
Alice and Bob have measured coincident results is sufficiently
small. The above inequalities come from the fact that in the case
of individual attacks the error on Eve's estimation of the final
bit $b$ is bound from below by a term proportional to
$|\braket{e_{++}} {e_{--}}|^N$ \cite{Navascues05}. Therefore,
Alice and Bob can establish a key if
\begin{equation}\label{Security}
    \frac{\epsilon_{AB}}{1 - \epsilon_{AB}} < |\braket{e_{++}}
    {e_{--}}|.
\end{equation}
In \cite{Navascues05} it was shown that any $1 \times 1$ NPPT
state fulfills the above inequality and thus any NPPT gaussian
state can be used to establish a secure key in front of individual
eavesdropper attacks.

Let us now present our results. Notice that since security relies
on the fact that Alice and Bob have better correlations than the
information the eavesdropper can learn about their state, perfect
correlation is not a requirement to establish a secure key. We
denote Alice's outputs by $x_{0A}$  and we calculate which are the
outputs Bob can accept so that the correlation established between
Alice and Bob outputs can be used to extract a secret bit.

We use the standard form of a bipartite $1 \times 1$ mode gaussian
state,
\begin{equation}
\gamma_{AB} = \begin{pmatrix}
    \lambda_A & 0 & c_x & 0\\
    0 & \lambda_A & 0 & -c_p\\
    c_x & 0 & \lambda_B & 0\\
    0 & -c_p & 0 & \lambda_B\\
    \end{pmatrix}
\end{equation}
with $\lambda_{A,B} \geq 0$, and $c_x \geq c_p \geq 0$ (we can shift
the DV to 0). The gaussian state is called symmetric if $\lambda_A
= \lambda_B = \lambda$ and fully symmetric if also $c_x = c_p$. We
shall deal with mixed symmetric states. The positivity
condition reads $(\lambda - c_x)(\lambda + c_p) \geq 1$, while the
entanglement NPPT condition is given by  $(\lambda - c_x)(\lambda - c_p)
< 1$. As in \cite{Navascues05}, we impose that the global state
including Eve is pure (she has access to all degrees of freedom
outside Alice an Bob) while the mixed symmetric state, shared by
Alice and Bob is just its reduction,
\begin{equation}
    \gamma_{ABE} =
    \begin{pmatrix}
     \gamma_{AB} & C\\
     C^T & \theta \gamma_{AB} \theta^T
    \end{pmatrix},
\end{equation}
\begin{equation}
    C = J_{AB} \sqrt{-(J_{AB} \gamma_{AB})^2 - \id_2} \, \theta_{AB} =
    \begin{pmatrix}
     0 & -\textsc{X} & 0 & -\textsc{Y}\\
     -\textsc{X} & 0 & -\textsc{Y} & 0\\
     0 & -\textsc{Y} & 0 & -\textsc{X}\\
     -\textsc{Y} & 0 & -\textsc{X} & 0\\
    \end{pmatrix},
\end{equation}
\begin{equation}
    \theta_{AB} = \theta_A \oplus \theta_B, \quad \quad J_{AB} = J_A
    \oplus J_B,
\end{equation}
where
$$\textsc{X}=\frac{\sqrt{a+b} + \sqrt{a-b}}{2},$$
$$\textsc{Y}=\frac{\sqrt{a+b} - \sqrt{a-b}}{2},$$
and $a = \lambda^2 -c_x c_p - 1$, $b = \lambda(c_x - c_p)$.

Performing a measurement with uncertainty $\sigma$, the
probability that Alice finds $\pm |x_{0A}|$ while Bob finds $\pm
|x_{0B}|$, is given by the overlap between the state of Alice and
Bob, $\hat\rho_{AB}$, and a pure product state $\hat \rho_{A,i}
\otimes \hat \rho_{B,j}$ (with $i,j=0,1$) of gaussians centered at
$\pm |x_{0A}|(\pm |x_{0B}|)$ respectively with $\sigma$ width
(notice $\hat\rho_{A,0} \equiv \ketbra{+|x_{0A}|}{+|x_{0A}|}$). We
use here the Hilbert-Schmidt fidelity which leads to:
\begin{equation}\label{Prob00}
    \begin{split}
     p(0,0) &= p(1,1) = {\rm tr}[\hat \rho_{AB} (\hat \rho_{A,0}
     \otimes \hat \rho_{B,0})] =\\
     &= (2\pi)^4 \int d^4
     \boldsymbol{\zeta}_{AB} \, \mathcal{W}_{\rho_{AB}}
     (\boldsymbol{\zeta}_{AB}) \mathcal{W}_{\rho_{A,0} \otimes
     \rho_{B,0}} (\boldsymbol{\zeta}_{AB}) =\\
     &= K(\sigma) \exp \left( \frac{2|x_{0A}| |x_{0B}| c_x - (\lambda
     + \sigma^2)(x_{0A}^2 + x_{0B}^2)}{(\lambda + \sigma^2)^2 -
     c_x^2} \right),
    \end{split}
\end{equation}
for the probability that their symbols do coincide and,
\begin{equation}\label{Prob01}
    p(0,1) = p(1,0) = K(\sigma) \exp \left( \frac{-2|x_{0A}| |x_{0B}|
    c_x - (\lambda + \sigma^2)(x_{0A}^2 + x_{0B}^2)}{(\lambda +
    \sigma^2)^2 - c_x^2} \right),
\end{equation}
for the probability that they do not coincide, where
\begin{equation}
    K(\sigma) = \frac{4\sigma^2}{\sqrt{(\lambda + \sigma^2)^2 -
    c_x^2}\sqrt{(\lambda \sigma^2 + 1)^2 - c_p^2 \sigma^4}}.
\end{equation}
The error probability for $\sigma \rightarrow 0$ reads
\begin{equation}\label{Errab}
    \epsilon_{AB} = \lim_{\sigma \to 0} \frac{\sum_{i \neq
    j}p\,(i,j)}{\sum_{i,j} p\,(i,j)} = \frac{1}{1 + \exp \left(
    \frac{4c_x |x_{0A}||x_{0B}|}{\lambda^2 - c_x^2} \right)}.
\end{equation}
Let us calculate the state of Eve $\ket{e_{\pm \pm}}$ after Alice
has projected onto $\ket{\pm |x_{0A}|}$ and Bob onto $\ket{\pm
|x_{0B}|}$:

\begin{equation}
    \gamma_{++} =  \gamma_{--} =
    \begin{pmatrix}
     \gamma_{x} & 0\\
     0 & \gamma_{x}^{-1}
    \end{pmatrix}, \quad \quad \gamma_{x} =
    \begin{pmatrix}
     \lambda & c_x\\
     c_x & \lambda
    \end{pmatrix},
\end{equation}

\begin{equation}
    d_{\pm\pm} = \mp
    \begin{pmatrix}
     0\\
     0\\
     A \delta x_0 - B \Delta x_0\\
     A \delta x_0 + B \Delta x_0
    \end{pmatrix},
\end{equation}
where $A = \frac{\sqrt{a+b}}{\lambda + c_x}$, $B
=\frac{\sqrt{a-b}}{\lambda - c_x}$, $\Delta x_0 = |x_{0B}| -
|x_{0A}|$ and $\delta x_0 = |x_{0B}| + |x_{0A}|$. The overlap
between the two states of Eve is given by:
\begin{multline}\label{Eveov}
    |\braket{e_{++}}{e_{--}}|^2 = \exp \Bigg(\frac{-4}{\lambda^2 -
    c_x^2} \Bigg[ \left( \frac{x_{0A}^2 + x_{0B}^2}{2} \right)
    (\lambda^2 - c_x^2-1)\lambda + \\ + |x_{0A}| |x_{0B}| \left(
    c_x - c_p(\lambda^2-c_x^2) \right) \Bigg] \Bigg).
\end{multline}
Substituting Eqs. \eqref{Errab} and \eqref{Eveov} into
\eqref{Security} one can check, after some algebra, that the
inequality \eqref{Security} reduces to:
\begin{equation}\label{SecurityIneq}
    \left(\frac{x_{0A}^2 + x_{0B}^2}{2}\right)(\lambda^2 -
    c_x^2 - 1)\lambda + |x_{0A}| |x_{0B}| \left( -c_x - c_p
    (\lambda^2 - c_x^2) \right) < 0.
\end{equation}
Notice that condition \eqref{SecurityIneq} imposes both,
restrictions on the parameters defining the state ($\lambda, c_x,
c_p$), and on the outcomes  of the  measurements ($x_{0A},
x_{0B}$).  The constraints on the state parameters are equivalent
to demand that the state is NPPT and satisfies
\begin{equation}\label{Constrain}
    (\lambda - c_x)(\lambda + c_x) \geq 1.
\end{equation}
Nevertheless, as $c_x \ge c_p$, any positive state fulfills this
condition. Hence for any NPPT symmetric state, there exists, for a
given $x_{0A}$, a range of values of $x_{0B}$ such that secret
bits can be extracted (Eq. \eqref{Security} is fulfilled). This
range is given by
\begin{equation}
    \Delta x_0 = |x_{0B}| - |x_{0A}| \in {\mathfrak D}_\alpha =
    \left[ \frac{2}{-\sqrt \alpha-1} , \frac{2}{\sqrt \alpha-1}
    \right] |x_{0A}|,
\end{equation}
where
\begin{equation}
    \alpha = \left( \frac{c_x -
    \lambda}{c_x + \lambda} \right) \left[ \frac{1 - (\lambda +
    c_x)(\lambda + c_p)}{1- (\lambda - c_x)(\lambda - c_p)}
    \right].
\end{equation}
After Alice communicates $|x_{0A}|$ to Bob, he will accept only
measurement outputs within the above interval. The interval
$\Delta x_0$ is well defined if $\alpha \geq 1$, which equivals to
fulfill Eq. \eqref{Constrain}. Notice also that the interval is
not symmetric around $|x_{0A}|$ because the probabilities
calculated in Eqs. \eqref{Prob00} and \eqref{Prob01} do depend on
this value in a non symmetric way. The length of the interval of
valid measurements outputs for Bob is given by
\begin{equation}
    D_\alpha = \frac{4 \sqrt \alpha}{\alpha-1} |x_{0A}|.
\end{equation}
It can be observed that maximal $D_\alpha\rightarrow\infty$
($\alpha=1$) corresponds to the case when Alice and Bob share a
pure state (Eve is disentangled from the system) and thus
condition \eqref{Security} is always fulfilled.  On the other
hand, any mixed NPPT symmetric state ($\alpha > 1$) admits a
finite $D_\alpha$. This ensures a {\it finite} efficiency on
establishing a secure secret key in front of individual attacks.

If we assume that Eve performs more powerful attacks, namely
finite coherent attacks, then security is only guaranteed if
\cite{Navascues05}:
\begin{equation}\label{Security2}
    \frac{\epsilon_{AB}}{1 - \epsilon_{AB}} < |\braket{e_{++}}
    {e_{--}}|^2.
\end{equation}
This condition is more restrictive than \eqref{Security}. With a
similar calculation as before we obtain that now security is not
guaranteed for all mixed entangled symmetric NPPT states, but only
for those that also satisfy:
\begin{equation}\label{Constrain2}
    \lambda - (\lambda+c_x)(\lambda-c_x)(\lambda-c_p) > 0.
\end{equation}
For such states, and given a measurement result $x_{0A}$ of Alice,
Bob will only accept outputs within the range:
\begin{equation}
    \Delta x_0 = |x_{0B}| - |x_{0A}| \in {\mathfrak D}_\beta =
    \left[ \frac{2}{-\sqrt \beta-1} , \frac{2}{\sqrt \beta-1}
    \right] |x_{0A}|,
\end{equation}
where
\begin{equation}
    \beta = \frac{2\lambda(\lambda^2-c_x^2-1)}{\lambda-
    (\lambda+c_x)(\lambda-c_x)(\lambda-c_p)} \geq 1.
\end{equation}
Eqs. \eqref{Constrain} and \eqref{Constrain2} already guarantee
that $\beta \geq 1$.

Let us now focus on the efficiency issue. We define the efficiency
$E(\gamma_{AB})$ of the protocol for a given state $\gamma_{AB}$,
as the average probability of obtaining a classically correlated
bit. Explicitly,
\begin{equation}\label{protocolefficiency}
    E(\gamma_{AB}) = \int_{\Delta x_0  \in {\mathfrak D}} dx_{0A} dx_{0B}
    (1-\epsilon_{AB}) {\rm tr} (\hat\rho_{AB} \ketbra{x_{0A},
    x_{0B}}{x_{0A},x_{0B}}).
\end{equation}
The marginal distribution in phase-space is easily computed by
integrating the corresponding Wigner function in momentum space
\cite{Lee95}:
\begin{equation}
    \begin{split}
     {\rm tr} (\hat\rho_{AB} \ketbra{x_{0A},x_{0B}}{x_{0A},x_{0B}})
     &= \int \int dp_A dp_B \mathcal{W}_{\rho_{AB}} (\boldsymbol
     {\zeta}_{AB}) =\\
     &= \frac{\exp \left( \frac{2 c_x x_{0A} x_{0B}
     - \lambda (x_{0A}^2+x_{0B}^2)}{\lambda^2-c_x^2} \right) }
     {\pi \sqrt{\lambda^2-c_x^2}},
    \end{split}
\end{equation}
but the final expression of \eqref{protocolefficiency} has to be
calculated numerically. Note that if Alice and Bob share as a
resource $N$ identical states (NPPT states for individual attacks,
and NPPT states fulfilling \eqref{Constrain2} for finite coherent
attacks), the number of classically correlated bits that can be
extracted from them is $\sim~N \times E(\gamma_{AB})$. The
efficiency \eqref{protocolefficiency} increases with increasing
$D$ and decreasing $\epsilon_{AB}$. In particular, for the
protocol given in \cite{Navascues05}, $D=0$, and therefore
$E(\gamma_{AB})=0$ for any state.

We investigate now the dependence of $E(\gamma_{AB})$ on the
entanglement of the NPPT mixed symmetric state used for the
protocol as well as on the purity of the state. As a measure of
the entanglement between Alice and Bob we compute the logarithmic
negativity
\begin{equation}
    {\rm LN} (\boldsymbol{\gamma}_{AB}) = \log_2 \left( \frac{1}
    {\sqrt{(\lambda-c_x)(\lambda-c_p)}} \right) > 0.
\end{equation}
\begin{figure}
    \centering
    \includegraphics[width=10cm]{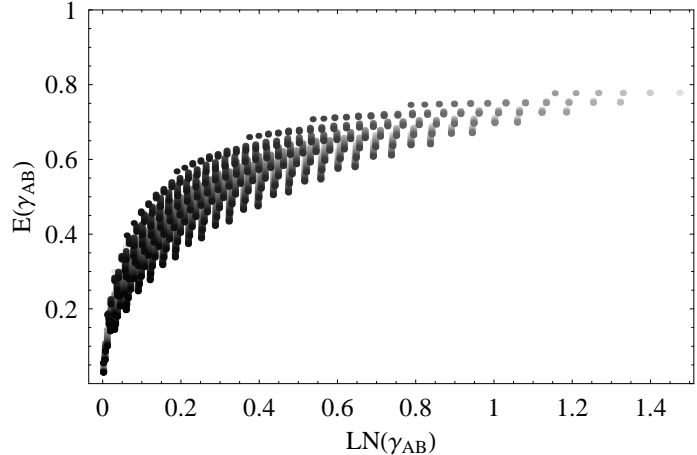}
    \caption{Protocol efficiency (quantified by $E(\gamma_{AB})$)
    versus the entanglement measured by logarithmic negativity
    ${\rm LN}(\gamma_{AB}$). The shading from black to white
    corresponds to purities from zero to one.}\label{fig2}
\end{figure}
In Fig. \ref{fig2}, we display the efficiency of the protocol
(assuming individual attacks) versus entanglement shared between
Alice and Bob for different states $\gamma_{AB}$. There is not a
one-to-one correspondence between $E(\gamma_{AB})$ and
entanglement, since states with the same entanglement can have
different purity, which can lead to different efficiency. This is
so because there are two favorable scenarios to fulfill
\eqref{Security}. The first one is to demand large correlations so
that the relative error $\epsilon_{AB}$ of Alice and Bob is small.
The second scenario happens when Alice and Bob share a state with
high purity, {\em i.e.}, Eve is very disentangled. In this case,
independently of the error $\epsilon_{AB}$, \eqref{Security} can
be fulfilled more easily.

Despite the fact that efficiency generally increases with
increasing entanglement, this enhancement, as depicted in the
figure, is a complex function of the parameters involved.
Nevertheless, one can see that there exist an entanglement
threshold (around ${\rm LN}(\gamma_{AB}) \simeq 0.2$) below which
the protocol efficiency diminishes drastically no matter how mixed
are the states shared between Alice and Bob.

It is also illustrative to examine the dependence of $\alpha$
(which determines the interval length $D_\alpha$) on the
entanglement of the states shared by Alice and Bob.
\begin{figure}
    \centering
    \includegraphics[width=10cm]{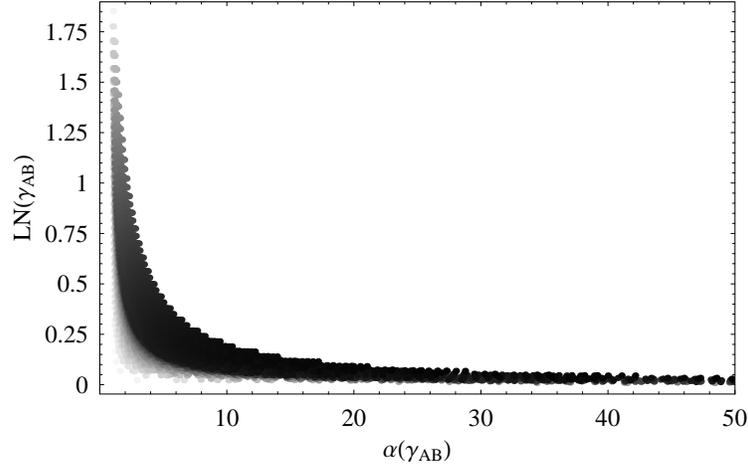}
    \caption{Entanglement of the states shared between Alice and Bob
    measured in terms of the logarithmic negativity ${\rm
    LN}(\gamma_{AB})$ versus the parameter $\alpha
    (\gamma_{AB})$ under individual attacks. The shading from black
    to white corresponds to purities from zero to one.}\label{fig1}
\end{figure}
In Fig. \ref{fig1} we plot the logarithmic negativity of a given
state versus the parameter $\alpha$. States with the same
entanglement but different purity are associated to quite
different values of $\alpha$. Nevertheless states with high
entanglement permit a large interval length (small $\alpha$) and,
thus, high efficiency.

In both, Fig. \ref{fig2} and Fig. \ref{fig1}, we have observed
that states with different entanglement give the same efficiency.
However it is important to point out that to extract the key's
bits, classical advantage distillation \cite{Maurer93} stills
needs to be performed. The efficiency of Maurer's protocol,
strongly increases with decreasing $\epsilon_{AB}$, and,
therefore, the states with higher entanglement will provide a
higher key rate.

\section{Conclusions}\label{Conclusions}

Efficiency is a key issue for any experimental implementation of
quantum cryptography since available resources are not unlimited.
Here, we have shown that the sharing of entangled gaussian
variables and the use of only gaussian operations permits
efficient quantum key distribution against individual and finite
coherent attacks. All mixed NPPT symmetric states can be used to
extract secret bits under individual attacks whereas under finite
coherent attacks and additional condition has to be fulfilled. We
have introduced a figure of merit (the efficiency $E$) to quantify
the number of classical correlated bits that can be use to distill
a key from a sample of $N$ entangled states. We have observed that
this quantity grows with the entanglement shared between Alice and
Bob. This relation it is not one-to-one due to the fact that
states with less entanglement but purer (Eve more disentangled)
can be equally efficient. Nevertheless as we have pointed out,
these states would be, inefficient in the distillation of the key.
Finally, we would like to remark that our study is not restricted
to quantum key distribution protocols, but can be extended to any
other protocol that uses as a resource entangled continuous
variable states to establish a set of classically correlated bits
between distant parties \cite{rodo2}.

{\it Acknowledgments} -- We thank A. Ac{\'\i}n, A. Monras, and J. Bae
for discussions. We acknowledge support from ESF PESC QUDEDIS, MEC
(Spanish Government) under contracts EX2005-0830, AP2005-0595,
CIRIT (Catalan Government) under contracts CSG-00185,
FIS2005-01369 and Consolider-Ingenio 2010 CSD2006-0019.

\end{document}